\documentstyle[12pt]{article}

\thicklines                         
\def\cs{{\cal S}}

\def\d#1/d#2{ {\partial #1\over\partial #2} }
\newcount\sectno

\def\sect{\global\advance\sectno by1 \the\sectno }
\def\pdr{\partial}

\def\al{\alpha}
\def\be{\beta}
\def\lm{\lambda}
\def\dag{\dagger}
\def\om{\omega}
\def\ga{\gamma}
\def\Ga{\Gamma}
\def\de{\delta}
\def\veps{\varepsilon}
\def\eps{\epsilon}
\def\bk{\bf k}
\def\half{{1\over 2}}

\def\NON{\nonumber}
\newcount\eqnumber
\def\beq{\begin{eqnarray}}
\def\eeq{\end{eqnarray}}
\def\n{\global\advance \eqnumber by 1\eqno(\the\eqnumber)}
\def\puteqno{\global\advance \eqnumber by 1 (\the\eqnumber)}
\def\beqs{\begin{eqnarray}}
\def\eeqs{\end{eqnarray}}
\newcount\refno\refno=0  
\def\ifundefined#1{\expandafter\ifx\csname #1\endcsname\relax}

\def\[#1]{
\ifundefined{#1}\advance\refno by
1\expandafter\edef\csname#1\endcsname{\the\refno}\fi[\csname #1\endcsname]}

\def\refis#1{\noindent\csname #1\endcsname. }

\def\label#1{
\ifundefined{#1}
\expandafter\edef\csname #1\endcsname{\the\eqnumber}
\else\message{label #1 already in use}
\fi{}}
\def\(#1){(\csname #1\endcsname)}
\def\eqn#1{(\csname #1\endcsname)}

\begin{document}
\begin{titlepage}

\hskip4.2in CU-TP-615

\hskip4.1in OCTOBER-93

\hskip4.1in hep-th/9311050
\vskip.5in

\begin{center}

 {\LARGE\bf Anomaly cancellation in 2+1 dimensions }

\vskip.1in

 {\LARGE\bf in the presence of a domainwall mass}

\vskip.4in

{\large  Shailesh Chandrasekharan$^\dag$} \\
{\large Department of Physics} \\
{\large Columbia University, New York, N.Y.10027 } \\
\date{\small 6 October 1993}

\end{center}

\vskip.4in

\begin{abstract}

	A Fermion in 2+1 dimensions, with a mass function which depends
on one spatial coordinate and passes through a zero ( a domain wall mass), is
considered. In this model, originally proposed by Callan and Harvey, the gauge
variation of the effective gauge action mainly consists of two terms.
One comes from the induced
Chern-Simons term and the other from the chiral fermions, bound to the
1+1 dimensional wall, and they are expected to cancel each other. Though
there exist arguments in favour of this, based on the possible forms of the
effective action valid far from the wall and some facts about theories of
chiral fermions in 1+1 dimensions, a complete calculation is lacking. In this
paper we present an explicit calculation of this cancellation at one loop
valid even close to the wall.
We show that, integrating out the``massive'' modes of the theory does produce
the Chern-Simons term, as appreciated previously. In addition we show that
it generates a term that softens the high energy behaviour of the 1+1
dimensional effective chiral theory thereby resolving an ambiguity present in
a general 1+1 dimensional theory.
\end{abstract}

\vskip.3in

\noindent --------------------------------

\noindent $^\dag$ email: sch@cuphyf.phys.columbia.edu

\end{titlepage}

\newpage

\section{Introduction}

	It was understood sometime ago that there exist intimate connections
between the Chern-Simons term in an odd dimensional space-time and the chiral
anomaly in one lower dimension. After such a connection was understood,
Callan and
Harvey \cite{call} proposed a model in which the connection was physically
realised. They considered a three dimensional fermion with a domain
wall mass (a mass term that depends on one space coordinate, passes
through zero at the origin and goes to a constant with opposite signs at
plus and minus infinity) coupled to a gauge theory. Since there
are no anomalies in the continuous symmetries in odd dimension, the theory
must be gauge invariant. However, in the theory with a domain wall mass,
one can show, as we will see later, that there exist effectively
two-dimensional massless chiral fermions attached to the domain wall. The
resulting two dimensional chiral theory should have an anomaly
in the gauge current. However, since the whole theory has no anomaly there
must be
yet another contribution to the current in the whole theory which will cancel
the chiral anomaly. It was found that there indeed exist currents on either
side of the wall which flow into or away from the wall depending on the
sign of the anomaly on the wall. This current can be approximately calculated
away from the wall using methods of Goldstone and Wilczek \cite{gold}. We will
call these currents Goldstone-Wilczek currents.

However, when one investigates the Goldstone-Wilczek currents flowing from the
third dimension into the wall and thus accounts for the charge appearance on
the wall(the person on the wall considers the charge appearing as an anomaly),
one encounters difficulties. In an abelian theory, for example the charge
appearing on the wall is twice as much as that predicted from an anomaly
in an exclusively 1+1 dimensional theory\cite{nacul}. In a non-abelian theory
the problem is more evident. Here, the anomaly in the two dimensional chiral
theory is necessarily gauge non-covariant.This non-covariant form is required
by the Wess-Zumino \cite{wess} consistency conditions obeyed by the usual
definition of the current. Hence the anomaly in this current is also referred
to as a
{\it consistent anomaly}. On the other hand the Goldstone-Wilczek current
accounts
for the anomaly on the wall which is gauge covariant in its form. This form of
the anomaly, which is gauge covariant in its form is also referred to as the
{\it covariant anomaly}. Thus this Goldstone-Wilczek current alone cannot
completely cancel the consistent anomaly on the wall, as one is gauge
covariant and the other is not.

			When Bardeen and Zumino discuss consistent and
covariant anomalies \cite{bard} they show how an addition of an extra term to
the consistent current can make the anomaly covariant in its form. Thus it
seems that there must exist an extra piece of current on the wall, that
arises naturally and which makes the anomaly in the effective 1+1 dimensional
theory covariant in its form. This term cannot be obtained from the
lagrangian of an exclusively
1+1 dimensional theory as the consistency conditions would not allow its
presence.On the other hand in our model this extra piece of the current can
be induced by the effects of the extra dimension. This problem was addressed
by
Naculich\cite{nacul} in which he suggests how a particular form of the
Chern-Simons term in the 2+1 dimensional effective action (produced when you
integrate out the massive fermion modes of the theory) can induce this extra
piece. In fact, this particular form of the Chern-Simons
term was originally suggested by Callan and Harvey \cite{call}. However there
is no complete derivation for this Chern-Simons term in the effective action,
which is valid arbitrarily close to the wall. This is not
satisfactory because, the calculations suggested to extract the effective
Chern-Simons term are valid only far from the wall, on the other hand the
actual questions, which are at issue here, are related to terms induced on the
wall. Also one wonders, if there is one extra effect on the wall, other than
that of a simple 1+1 dimensional massless chiral theory,
there may be others that are hidden and unclear,until a complete
calculation of the effective action is done.

	A more clean and simple view of this anomaly cancellation comes from
considering an effective action in terms of the gauge fields after
integrating out the fermions. This effective action must be gauge invariant.
This was the original way Callan and Harvey \cite{call} analysed the problem.
In their paper \cite{call} they argue that integrating out the massive
fermion modes (the modes that are not chiral and do not live on the wall)
produces a Chern-Simons term in the effective action. They show how the
gauge variation of this term can cancel the variation coming from the
remaining 1+1 dimensional effective chiral theory on the wall. This
cancellation, when examined in terms of the currents, motivates the above
discussion of covariant and consistent anomalies considered first by
Naculich \cite{nacul}. Hence, even in this
simpler view, one needs a derivation of the suggested Chern-Simons term
in the effective action that is valid close to the wall. Further one must show
that, the resulting 1+1 dimensional theory can be treated as a naive 1+1
dimensional massless chiral theory with no other effects induced.

    These questions are of particular interest in the context of the recent
proposal, by Kaplan, to solve the doubling problem on the lattice \cite{kap}.
The basic model used in this new proposal is the same as the one proposed by
Callan and Harvey. In this model one is looking for a theory of massless
chiral fermions on the wall. To make the
final theory exclusively live on the wall it is
important that the massive modes have very little effect on the wall, as the
massive modes are presumed to decouple from the theory on the wall. So that
the issues discussed above are important in this context.

	 Taking all this into account it seems quite important
to understand the structure of the cancellation of the anomaly in this model
with a domain wall mass. An explicit calculation would clarify the effects of
the massive modes in this cancellation. Also the calculations given in
references \cite{call} and \cite{nacul} deal mainly with
axionic strings apart from mentioning the applicability to domain walls. In
view of the recent interest in the domain wall problem \cite{kap}
we think it makes sense to write down some results explicitly, valid for
the domain wall case along with some proofs for the previously suggested
results.

				 In this paper we study the model in which
fermions interact with a domain wall, which is a smooth function of one space
direction,and
couple to an abelian gauge field. In section $2$ we study the eigenstates for
the free dirac operator with this domain wall mass and show how the
eigenstates of the theory change as the steepness of the mass function
changes. We actually find that, as the mass function becomes smooth, the number
of states bound to the wall increases, though only one is chiral. We
then pick a particular mass function, which turns out to be easy to analyse
and which has only one (chiral)bound state, and derive the complete set of
eigenstates for the free dirac operator with this choice of the mass
function. In section $3$ we find the free propagator for
this theory including the effects of the space-dependent mass term using the
exact eigenstates derived in section $2$ for the particular choice of the
mass function. As computations are much easier in Euclidean space, we continue
our results to Euclidean space and obtain a closed form expression for the
propagator in Euclidean configuration space.  In section $4$ we integrate out
the fermion fields using this Euclidean-space propagator and treat the
gauge coupling perturbatively, to compute the one loop effective action for
the gauge fields.
When we look at only the terms potentially contributing to the anomaly, (i,e..
the terms that contain the completely antisymmetric tensor) we find that, in
the low energy limit the effective gauge action consists of two terms. The
first
is the old Chern-Simons term, as suggested by Naculich \cite{nacul}, but now
without any assumptions and valid arbitrarily close to the wall. The second
term is the chiral term, which has contributions not only from the chiral
bound states but also from fully three-dimensional massive states. The extra
contribution from the massive states acts to ``regulate'' the chiral term.
Thus we find that the chiral anomaly is generated not by a
potentially singular term but by a well regulated term. The high energy
characteristic of the chiral fermions on the wall is ``softer'' than the  usual
chiral fermions in two dimensions. Finally we show how,because of these
effects, the gauge variation of the Chern-Simons term and the chiral term in
the effective gauge action cancel each other explicitly. Before concluding we
show how this cancellation can be viewed in terms of the currents in section
$5$.

\section{States of the Theory}

	We start with a theory defined in Minkowski space to give the theory
a physical setting, and to be able to analyse the physical states of the
theory. The theory can be given in terms of the action
\beq
	\cs = \int  d^3z\  \overline\Psi\{i\ga^\mu(\pdr_\mu -ieA_\mu) +m(s)\}
\Psi
\eeq
where the $\ga^\mu$'s are the dirac matrices , which obey the
anticommutation relations
\beq
	\{ \ga^\mu,\ga^\nu \} = g^{\mu\nu} ; \mu,\nu = 0,1,2
\eeq
and $\overline\Psi= \Psi^\dag\ga^0$.
We assume that $\mu = 0$ is the ``time'', $\mu = 1,2$ the space direction.
So that the metric would be $g^{00} = 1, g^{11} = -1$ and $g^{22} = -1$.
The coordinates of the space-time are labeled by $z^\mu = (t,x,s)$
The mass depends on the second space direction, labeled by $s$.
We want $m(s)$ to be a function with a domain wall shape,i.e.
\beq
	m(s) = \left\{ \matrix{
			m_0  &  s \rightarrow +\infty \cr
			-m_0  &  s \rightarrow -\infty \cr
			0  & s = 0 \cr}
		\right.
\eeq
We first want to solve for the states of the theory to see the effects
of the mass function. In order to proceed further we assume a specific
form of the mass function which is sufficiently general to be able to
study its various limiting forms. We choose
\beq
	m(s) = m_0\hbox{tanh}\mu_0s
\eeq
To be concrete we choose a $chiral$ representation for the Dirac matrices
defined by,
\beq
\left.	\ga^0= \left( \matrix{0 & 1 \cr
				1 & 0 \cr} \right)
;\ 	\ga^1= \left( \matrix{0 & 1 \cr
				-1 & 0 \cr} \right)
;\ 	\ga^2= \left( \matrix{-i & 0 \cr
				0 & i \cr} \right)
\right.
\eeq
Having fixed the theory lets ask what the eigenstates of the theory
are? We will treat the gauge coupling as a perturbation, and think of
the remaining theory as a {\it free} theory and solve  the equations
of motion. However, keeping in mind that we would like to solve for the
propagator of the theory too, we will try to solve the following eigenvalue
equation.
\beq
	\ga^0\ \{i\ga^\mu\pdr_\mu + m(s)\}\Psi_\lm = \lm\Psi_\lm
\eeq
or in the matrix form this can be written as
\beq
\left.	\left[ \matrix{i\pdr_t-i\pdr_x & -\pdr_s + m(s) \cr
			\pdr_s + m(s) & i\pdr_t + i\pdr_x \cr}
	\right]
	\left[ \matrix{\psi_1 \cr
			\psi_2 \cr}
	\right]
= \lm 	\left[ \matrix{\psi_1 \cr
			\psi_2 \cr}
	\right]
\right.
\eeq

Clearly the eigen-solutions for $\lm$ = 0 will also be the solutions for
the equations of motion. The propagator for the theory, $S(z,z^\prime)$
can then be constructed from the eigen-solutions above by
\beq
	S(z,z^\prime) = \sum_{\lm} {\Psi(z)\overline\Psi(z^\prime)\over\lm}
\eeq
 When we construct the propagator we must remember to take the feynman
prescription that can be derived using the usual canonical commutation
relations. However lets first solve for the eigenvalues $\lm$ and the
corresponding eigenfunctions $\Psi_\lm$.

To do this, first observe that the eigen functions can be written in the
form $\Psi_\lm = \Phi_{\lm,\bk}(s)e^{-ik_0t}e^{-ik_1x}$, where we
characterize the eigenvalues $\lm$ by $\bk = (k_0,k_1,k_2)$. Actually we can
look at the asymptotic behaviour of eq. (7)  and  fix the eigenvalues. We
then find that for a given $k_0,k_1,k_2$ we get two values of $\lm$ given by
\beq
	\lm_\pm = k_0 \pm  \om_k
;\ \  \om_k = (k_1^2+k_2^2+m_0^2)^{\half}
\eeq
Using these facts
one can write the differential equations obeyed by the two components of
$\Phi_{\lm,\bk}, \phi_1$ and $\phi_2$, as
\beq
\left.	\left[ \matrix{k_0-k_1 & -\pdr_s + m(s) \cr
			\pdr_s + m(s) & k_0 + k_1 \cr}
	\right]
	\left[ \matrix{\phi_1 \cr
			\phi_2 \cr}
	\right]
= \lm_\pm \left[ \matrix{\phi_1 \cr
			\phi_2 \cr}
	\right]
\right.
\eeq
One can rewrite the above two coupled equations for $\phi_1$ and $\phi_2$
as decoupled second order equations given below.
\beq
\left.
\matrix{	[{\pdr^2\over\pdr s^2} + [k_2^2 + m_0^2 - m^2(s) +
			{\pdr m(s)\over\pdr s}]\phi_1 = 0 \cr
\cr
	[{\pdr^2\over\pdr s^2} + [k_2^2 + m_0^2 - m^2(s) -
			{\pdr m(s)\over\pdr s}]\phi_2 = 0 \cr
}
\right.
\eeq
If we solve for $\phi_1$ or $\phi_2$ we can substitute it in
(10) and obtain the other by solving a simple algebraic equation.
Substituting $m(s)= m_0\hbox{tanh}\mu_0s$ in (11) we get
\beq
\left.
\matrix{
	[{\pdr^2\over\pdr s^2} + k_2^2 + \al(\al+1)
			{\mu_0^2\over \hbox{cosh}^2\mu_0s}]\phi_1 = 0 \cr
\cr
	[{\pdr^2\over\pdr s^2} + k_2^2 + \al(\al-1)
			{\mu_0^2\over \hbox{cosh}^2\mu_0s}]\phi_2 = 0 \cr
}
\right.
\eeq
where $\al = {m_0\over\mu_0}$.

We wish to solve (12). In fact these two equations describe quantum
mechanical scattering off a {\it modified P\"{o}schl Teller potential} and the
energy eigenvalues and functions are known, see for example \cite{flug}.
The potential can be written generically as $ \be(\be+1){\mu_0^2\over
 \hbox{cosh}^2\mu_0s}$, and which is known in the literature as the
{\it modified P\"{o}schl Teller potential}. It is clear from (12) that the
only difference between $\phi_1$ and $\phi_2$ is that the $\be = \al$ for
$\phi_1$ and $\be = \al -1$ for $\phi_2$. It is known that given $\be$, the
number of bound states equals the largest integer less than $\be + 1$. Clearly
$\phi_1$ has always one more bound state than $\phi_2$. This is the chiral
bound state since it can be shown that if this $\phi_1$ is substituted back in
(10) it gives $\phi_2 = 0$. This is the chiral state which is responsible for
the anomaly in [1]. However we also see here that as the mass function becomes
less steep, i.e. as ${m_0\over\mu_0}$ becomes large, the number of bound states
increases, though none of these are chiral except the one discussed above.
If $0 < {m_0\over\mu_0} \leq 1$ the only bound state is the chiral bound
state. As this is the case of most interest at present we will assume
$m_0 = \mu_0$, so that the mass function becomes $m_0\hbox{tanh}m_0s$.
The reason for doing this is that then the eigenstates of the theory are can
be written
in closed form. Substituting that $\al = {m_0\over\mu_0}=1$ in (12) we get
\beq
\left.
\matrix{
	[{\pdr^2\over\pdr s^2} + k_2^2 +
			{2\mu_0^2\over \hbox{cosh}^2\mu_0s}]\phi_1 = 0 \cr
\cr
	[{\pdr^2\over\pdr s^2} + k_2^2 ]\phi_2 = 0 \cr
}
\right.
\eeq
It is now evident how the choice of $\al$ as above simplifies things! We can
solve (13) for $\phi_2$ in the above equation and substitute it back in eq.(10)
to obtain $\phi_1$.  These will be the scattering eigenstates, given by
\beq
	\Phi_{\lm_\pm,\bk} = \left[ \matrix{ ik_2 + m(s) \cr
					\pm\om_k + k_1 \cr
					} \right] e^{-ik_2s}
\eeq
where $m(s) = m_0\hbox{tanh}m_0s$.
(We will hereafter assume $m(s) = m_0\hbox{tanh}m_0s$ wherever we use $m(s)$).
Note that when we substitute $\phi_2$ back in (10) we will in general get two
solutions for $\phi_1$ depending on the sign $\pm\lm$.
 This is explicitly shown in
(14). Note that at present the range of $k_2$ goes from $-\infty$ to $\infty$.
However we have not yet obtained all the eigenstates of the theory.
There exists one bound state for $\phi_1$ which can be obtained by solving
(13). This cannot be obtained from a solution of $\phi_2$ because for this
eigenstate $\phi_2 =0$. This bound state is given by
\beq
	\Phi_{\rm chiral} = \left[ \matrix{ \hbox{sech}m_0s \cr
					0 \cr
					} \right]
\eeq
Along with this we have the complete set of states of the theory though they
are not yet orthonormal. We can take linear
combinations and form an orthonormal basis for the theory which we find to be
\beq
\left.
\matrix{
	\Psi_{\bk,\lm_\pm,{\rm odd}} = & ({1\over 8\pi^3}
				{1\over \om_k (\om_k \pm k_1)})^\half
\left[ \matrix{ k_2\hbox{cos}k_2s - m(s)\hbox{sin}k_2s \cr
		-(\pm\om_k + k_1)\hbox{sin}k_2s \cr
		}
\right]e^{-ik_0t-ik_1x}   \cr
\cr
\cr
	\Psi_{\bk,\lm_\pm,{\rm even}} = & ({1\over 8\pi^3}
				{1\over \om_k(\om_k \pm k_1)})^\half
\left[ \matrix{ k_2\hbox{sin}k_2s + m(s)\hbox{cos}k_2s \cr
		(\pm\om_k + k_1)\hbox{cos}k_2s \cr
		}
\right]e^{-ik_0t-ik_1x}  \cr
\cr
\cr
	\Psi_{\bk,\lm_0,\rm chiral}\ \ = & \hskip .4in
({1\over4\pi^2}{m_0\over 2})^\half\left[ \matrix{ \hbox{sech}m_0s \cr
		0 \cr
		}
\right]e^{-ik_0t-ik_1x} \cr
}
\right.
\eeq
with eigen values, $\lm_\pm,\lm_\pm,\lm_0 = k_0-k_1$, respectively.

  The subscript $\bk,\lm,\alpha$ characterizes the
different eigenstates. $\lm$ refers to the eigen value, $\alpha$ refers to odd,
even or chiral, and $\bk$ refers to $(k_0,k_1,k_2)$. The allowed range of
$\bk$ is given by
\beq
	-\infty < k_0,k_1 < +\infty  \hbox{ and } 0 \leq k_2 < +\infty
\eeq
The asymmetry in the range for $k_2$ arises because we have made linear
combinations of $+k_2$ and $-k_2$ to form odd and even states.

The orthonormality can be tested by showing
 $\int \Psi^\dag_{\bk,\lm,\alpha}\Psi_{\bk^\prime,\lm^\prime,\alpha^\prime} =
    \de_{\alpha,\alpha^\prime}\de_{\lm,\lm^\prime}\de^3 (\bk - \bk^\prime)$
when $\alpha$ and $\alpha^\prime$ are not chiral. If both $\alpha$ and
$\alpha^\prime$ are chiral then the three dimensional delta function is
replaced by a two dimensional delta function in $k_0$ and $k_1$. The chiral
eigenstate is orthogonal to both, the odd and the even, eigenstates.
 Thus we have an explicit calculation for the eigenstates of the theory. We
can now go ahead and construct the propagator as described by (8).

\section{Propagator}

The construction of the propagator is straight forward though the derivation
of the explicit expression in closed form will be complicated. We use (8) to
obtain an integral representation in ``momentum'' space for the propagator.
After substituting and rearranging the terms and also extending the limits of
$k_2$ integral from $-\infty$ to $+\infty$ we get
\beq
S(z,z^\prime) = S_{\rm Chiral}(z,z^\prime) + S_{\rm massive}(z,z^\prime)
\eeq
where the chiral part is due to the chiral mode in the summation in (8) and
the massive part is due to the rest. The two terms are given by
\beqs
S_{chiral} & = & \half (1+i\ga^2){m_0\over 2}
		\ \hbox{sech}(m_0s) \hbox{sech}(m_0s^\prime)
		\int {d^2k\over (2\pi)^2}
		 \ {\ga^ak_a \over k_0^2 - k_1^2 + i\eps}
\NON \\
&   &
\hskip 1.5in \hbox{ x }e^{-ik_0(t-t^\prime)} e^{-ik_1(x-x^\prime)}
\eeqs
\beq
S_{massive} & = & \int {d^3k\over (2\pi)^3} \
		{[\ga^\mu k_\mu + M]\over k^2-m_0^2+i\eps}
	e^{-ik_0(t-t^\prime)} e^{-ik_1(x-x^\prime)}e^{-ik_2(s-s^\prime)}
\eeq
where $M$ is a matrix in spinor space given by
\vspace{1ex}
\beq
M = \left[ \matrix{ -m(s) & {k_0+k_1\over k_2^2+m_0^2}[m(s)m(s^\prime)
		+ ik_2(m(s^\prime)-m(s)) - m_0^2]  \cr
	 	0 & -m(s^\prime) \cr}
\right]
\eeq
 To make the notation clear
we will use Greek letters to run from 0 to 2 and Latin letters run from
0 to 1; so that $\ga^a k_a$ means $\ga^0 k_0 +\ga^1 k_1$ unlike
$\ga^\mu k_\mu$ which means $\ga^0 k_0 +\ga^1 k_1 +\ga^2 k_2$. The
above form of the propagator has a structure similar to the usual fermion
propagator in three dimensions except for the massless chiral term, which
reflects the chiral modes on the wall and the unusual mass matrix $M$. The
integral in the mass matrix $M$ is the only thing that
seems quite difficult to explicitly evaluate. Note that in three dimensions the
other integrals can be easily done!

At this stage we will analytically continue to the {\it Euclidean space} so
that we can explicitly evaluate the propagator and study the gauge
transformation  properties of the one loop effective action. This can be
most easily done in {\it Euclidean space} where things
are well defined. The continuation to {\it Euclidean space} means
\beq
  t = -i\tau ; \ k_0 \rightarrow ik_0;
\ \Ga^0 = \ga^0;\Ga^1=-i\ga^1;\Ga^2=-i\ga^2;
\eeq
where the $\Ga$'s are the gamma matrices in Euclidean space and $\tau$ is
the Euclidean time. After the explicit calculations are done, we find the
final result for the {\it Euclidean} propagator is
\begin{eqnarray}
S_E(z,z^\prime) & = &  -{\Ga^\mu\over 4\pi}{r_\mu\over r^3}(1+m_0r)e^{-m_0r}
+ {1\over 4\pi}{e^{-m_0r}\over r}\left [ \matrix{m(s) & 0 \cr
			0 & m(s^\prime) \cr} \right] \NON \\ \NON \\
&   &
 - {1\over 8\pi}(1-\Ga^2){\Ga^a\veps_a\over \veps^2}
{m_0e^{-m_0r}\over r} \NON \\
&   &
\hskip1in \hbox{ x }
[r\{ {m(s)\over m_0}{m(s^\prime)\over m_0}  - 1 \} + (s-s^\prime)
\{ {m(s^\prime)\over m_0} - {m(s)\over m_0} \}] \NON \\ \NON \\
&   &
 - {1\over 8\pi}(1-\Ga^2){\Ga^a\veps_a\over \veps^2}
m_0\hbox{sech}(m_0s)\hbox{sech}(m_0s^\prime)
\end{eqnarray}
\vskip.1in
\noindent where the various symbols are defined as
\vskip.05in
\beq
\left.
\matrix{
	r^\mu = (\tau-\tau^\prime,x-x^\prime,s-s^\prime); &
	\veps^a = (\tau-\tau^\prime,x-x^\prime); \cr
\cr
	\veps = \sqrt{(\tau-\tau^\prime)^2+(x-x^\prime)^2}; &
	r = \sqrt{\veps^2+(s-s^\prime)^2}; \cr
}
\right.
\eeq
\vskip.05in
The last term of the propagator can be easily identified with the {\it chiral}
propagator except for the sech$(m_0s)$sech$(m_0s^\prime)$ term which says that
these modes are bound to the wall. The first two terms combined become
essentially the three dimensional massive propagator with the mass term
modified
into a matrix. The third term seems to be new. It has the character of the
chiral term ( the last term) but at the same time is massive. To make this
explicit we rewrite (23) in a slightly different form by combining the third
and the fourth terms. After substituting $m(s) = m_0\hbox{tanh}m_0s$, we get
\beqs
S_E(z,z^\prime) & = &  -{\Ga^\mu\over 4\pi}{r_\mu\over r^3}(1+m_0r)e^{-m_0r}
+ {1\over 4\pi}{e^{-m_0r}\over r}\left [ \matrix{m_0\hbox{tanh}m_0s & 0 \cr
			0 & m_0\hbox{tanh}m_0s^\prime \cr} \right]
\NON \\ \NON \\
&   &
  - {1\over 8\pi}(1-\Ga^2){\Ga^a\veps_a\over \veps^2}
[1 - f(\veps,s-s^\prime)]
m_0\hbox{sech}(m_0s)\hbox{sech}(m_0s^\prime)
\eeqs
where the function $f(\veps,s-s^\prime)$ is given by
\beq
f(\veps,s-s^\prime) = {e^{-m_0r}\over r}[r\hbox{cosh}m_0(s-s^\prime)
+(s-s^\prime) \hbox{sinh}m_0(s-s^\prime)]
\eeq
We have written the third term in (23) in terms of the function $f$ so as
to unite it with the chiral term. Now the effects of this term are clearer.
It modifies the singularity structure of the massless chiral term. This
suggests that the massive modes ``regulate'' the chiral modes. This
also suggests that, when the theory is coupled to a gauge field, a
cancellation of the anomaly between the massive modes and the massless chiral
modes might be more involved. Hence we investigate the cancellation of the
anomaly in the next section.

\section{Anomaly cancellation at 1-loop}

	Having found the free propagator explicitly we can treat the gauge
coupling perturbatively and construct
the one-loop effective gauge action induced by integrating out the fermion
fields. Note that the effective action being
calculated here is the one-loop gauge action with no fermion fields. It is
still three dimensional though parts of it might look two dimensional
due to the presence of the chiral pieces that are non zero only on the
wall. This is important to remember as there are many other kinds of
effective action that can be considered, for example by integrating out
only the massive fermion fields and keeping the effective action dependent
on the chiral fields. This would be natural when one wants to study the
effective chiral fermion theory in the low energy limit. However we are not
doing this. We want to study the gauge invariance of the full theory and that
can be done by integrating out all the fermion fields, and studying the full
three-dimensional effective gauge theory. However we will treat the
fluctuations in the gauge fields as small compared to the mass of the massive
modes of the fermion fields in order to motivate a low energy effective gauge
theory. We will then
show that this low energy effective theory is gauge invariant. The actual
problem at issue here is the gauge invariance of this low energy effective
gauge theory, as there are non trivial low energy terms that are induced by
both the massless chiral fermion fields and also the massive fermion fields.

Having explained our motive lets look at the one-loop effective gauge action,
which is given by,
\beq
\cs_{eff}[A]
=\half \int d^3z\ d^3z^\prime A_\mu(z)V^{\mu\nu}(z,z^\prime)A_\nu(z^\prime)
\eeq
where
\beq
V^{\mu\nu} = tr[\Ga^\mu S_E(z,z^\prime)\Ga^\nu S_E(z^\prime,z)]
\eeq
The trace is over the spinor space.

The expression for $V^{\mu\nu}$ clearly
would be quite a long and complicated expression when $S_E(z,z^\prime)$ is
substituted, but as we will be interested finally in the $m_0\rightarrow\infty$
limit it makes sense to just to look at the expression for limit. Also we
will focus attention on the part of $V^{\mu\nu}$ which has either a two
dimensional $\eps^{ab}$ or a three dimensional  antisymmetric tensor
$\eps^{\mu\nu\rho}$. This is because the potential anomaly occurs in this part.
The appropriate expression for $V^{\mu\nu}_{\rm a}$ (where the label ${\rm a}$
denotes this potentially anomalous part), keeping only the nontrivial terms
that survive in the limit $\hbox{m}_0 \rightarrow \infty$, is found to be

\beqs
V^{\mu\nu}_{\rm a} & = &
{1\over 8\pi^2}
[-i\eps^{\mu\nu\rho}{r_\rho\over r^4}(1+m_0r)e^{-2m_0r}(m(s) + m(s^\prime))
\NON \\ \NON \\
&  &
- \ {m_0^2\over 2}{\veps^\mu\veps^{\ast\nu} +
			\veps^\nu\veps^{\ast\mu}\over\veps^4}
(1-f(\veps,s-s^\prime))^2\hbox{sech}^2(m_0s)\hbox{sech}^2(m_0s^\prime)]
\eeqs
where $\eps^{\mu\nu\rho}$ is the totally antisymmetric tensor with
$\eps^{012} = 1$ and $\veps^{\ast a} = i\eps^{ba}\veps_b$ is the
dual of $\veps^a$. Note that $\veps^a$ was defined in (24). Here
$\eps^{ab}$ is the antisymmetric tensor in two dimensions. Note also that
whenever
the two vector  $\veps^\mu$ is encountered, the component $\veps^2$ is
assumed to be zero.

As stated above it is assumed that $m_0\rightarrow \infty$ will be eventually
taken and only the relevant terms in this limit are given. The important point
to note is that though the function $f(\veps,s-s^\prime)$ comes from the
massive modes we cannot throw it away because in the limit
$\veps \rightarrow 0$
this function goes to 1 and hence contributes to the anomaly in a non-trivial
way. If the above expression for $V^{\mu\nu}_a$ is substituted in (28) and
some simplification is done we get
\beqs
\cs_{eff}^{\rm a} & = & \cs_{eff}^{\rm cs} + \cs_{eff}^{\rm chiral}
\NON \\ \NON \\
	  & = &  {-i\over 8\pi}\int d^3z {\rm sgn}(s)\eps^{\mu\rho\nu}
						A_\mu\pdr_\rho A_\nu
\ - \ {1\over 16\pi^2}\int d^3z d^3z^\prime
			A_a(z) \half{\veps^a\veps^{\ast b}
+ \veps^b\veps^{\ast a}\over\veps^4}A_b(z^\prime)
\NON \\ \NON \\
&   &
\hskip 1.0in
\hbox{  x  }m_0^2[1-f(\veps,s-s^\prime)]^2
		\hbox{sech}^2(m_0s)\hbox{sech}^2(m_0s^\prime)
\eeqs

	The first term is the Chern-Simons term which is mentioned in [1].
The limit of $m_0\rightarrow\infty$ turns $\tanh m_0s$ to ${\rm sgn}(s)$ which
is
the origin of the ${\rm sgn}(s)$ in (30). We have also used the limit
\beq
lim_{m_0\rightarrow\infty}  m_0{e^{-2m_0r}\over r^2} = 2\pi\delta({\bf r})
\eeq
Note also that we are analysing the term in the effective action which has
in it the antisymmetric tensor and hence the superscript $a$ in the action.
We must now show that $\delta \cs_{eff}^a$, the gauge variation, is zero.

	First lets consider the Chern-Simons term denoted by
$\cs_{eff}^{\rm cs}$. We then have
\beq
	\delta \cs_{eff}^{\rm cs} =
 {-i\over 8\pi}\int d^3z {\rm sgn}(s)\eps^{\mu\rho\nu}
					(\delta A_\mu\pdr_\rho A_\nu
\  + \ A_\mu\pdr_\rho \delta A_\nu)
\eeq
where $\delta A_\mu = \pdr_\mu \Theta$. Substituting this and also using
the antisymmetry of $\eps^{\mu\rho\nu}$ we get
\beqs
	\delta \cs_{eff}^{\rm cs} & = &
 {-i\over 8\pi}\int d^3z {\rm sgn}(s)\eps^{\mu\rho\nu} \pdr_\mu
						\Theta\pdr_\rho A_\nu
\NON \\ \NON \\
& = &
 {i\over 8\pi}\int d^3z 2\delta(s)\eps^{ab} \Theta\pdr_a A_b
\NON \\ \NON \\
& = &
 {i\over 4\pi}\int d^2z \Theta(z)[\eps^{ab}\pdr_a A_b]
\eeqs

	Now let us consider the second term, we shall call this the
{\it chiral} term. Note that there are factors of $m_0$ present in this
term because we cannot take the limit $m_0\rightarrow\infty$ before
the integration. Making a gauge variation as above we get
\beqs
\delta \cs_{eff}^{\rm chiral} & = & - {1\over 16\pi^2}\int d^3z d^3z^\prime
	m_0^2\hbox{sech}^2(m_0s)\hbox{sech}^2(m_0s^\prime)
\NON \\ \NON \\
&   &
(\pdr_a\Theta(z)A_b(z^\prime) + A_a(z)\pdr_b\Theta(z^\prime))
\hbox{ x } \half{\veps^a\veps^{\ast b} + \veps^b\veps^{\ast a}\over\veps^4}
[1-f(\veps,s-s^\prime)]^2
\NON \\ \NON \\
& = & {1\over 8\pi^2}\int d^3z d^3z^\prime
m_0^2\hbox{sech}^2(m_0s)\hbox{sech}^2(m_0s^\prime) \Theta(z) A_b(z^\prime)
\NON \\ \NON \\
&   &
\hskip 1.0in \hbox{ x }
\pdr_a(\half{\veps^a\veps^{\ast b} + \veps^b\veps^{\ast a}\over\veps^4}
[1-f(\veps,s-s^\prime)]^2)
\eeqs
Using the fact that the singularity as $\veps\rightarrow 0$ is not severe
due to the factor $[1-f(\veps,s-s^\prime)]^2$ we can compute the partial
derivative. We get the result
\beqs
\pdr_a({\veps^a\veps^{\ast b} + \veps^b\veps^{\ast a}\over\veps^4}
[1-f(\veps,s-s^\prime)]^2)
& = & {\veps^a\veps^{\ast b} + \veps^b\veps^{\ast a}\over\veps^4}
{\pdr\over\pdr\veps}([1-f(\veps,s-s^\prime)]^2){\veps_a\over\veps}
\NON \\ \NON \\
& = &  i\eps^{cb}{\veps^c\over\veps^3}
{\pdr\over\pdr\veps}([1-f(\veps,s-s^\prime)]^2)
\eeqs
where we have used the fact that $\veps^{\ast a}\veps_a = 0$ and the definition
of the dual $\veps^{\ast b} = i\eps^{cb}\veps^c$. Using the above result we
have
\beqs
\delta \cs_{eff}^{\rm chiral} & = & \ {i\over 16\pi^2}\int d^3z
d^3z^\prime m_0^2\hbox{sech}^2(m_0s)\hbox{sech}^2(m_0s^\prime)
\Theta(z)\eps^{cb} A_b(z^\prime)
\NON \\ \NON \\
&   &
\hskip 1.0in \hbox{ x }
{\veps^c\over\veps^3}
{\pdr\over\pdr\veps}([1-f(\veps,s-s^\prime)]^2)
\eeqs

Now we can use the property that $m_0$ is large and that contributions to the
integral come only from the region $z^\mu - z^{\prime \mu} = r^\mu$ tends to
zero. So we can expand $A_b(z^\prime)$ about $z^\prime = z$ and keep only the
terms that do not vanish in the $m_0\rightarrow\infty$, we get
\begin{eqnarray*}
\delta \cs_{eff}^{\rm chiral} & = & {i\over 16\pi^2}\int d^3z d^3r
m_0^2\hbox{sech}^2(m_0s)\hbox{sech}^2(m_0s^\prime)
\Theta(z)\eps^{cb}\pdr_a A_b(z)(-\veps^a)
\\
&   & \hskip 1.0in \hbox{ x }
{\veps^c\over\veps^3}
{\pdr\over\pdr\veps}([1-f(\veps,s-s^\prime)]^2)
\\
& = & -{i\over 32\pi^2}\int d^3z d^{s^\prime} d^2\veps
m_0^2\hbox{sech}^2(m_0s)\hbox{sech}^2(m_0s^\prime)
\Theta(z)\eps^{ab}\pdr_a A_b(z)
 \\
&   & \hskip 1.0in \hbox{ x }
{1\over\veps}
{\pdr\over\pdr\veps}([1-f(\veps,s-s^\prime)]^2)
\end{eqnarray*}
using $d^2\veps = 2\pi\veps d\veps$  we get
\beqs
\delta \cs_{eff}^{\rm chiral} & = & -{i\over 16\pi}\int d^3z ds^\prime
m_0^2\hbox{sech}^2(m_0s)\hbox{sech}^2(m_0s^\prime)
		\Theta(z)\eps^{ab}\pdr_a A_b(z)
\NON \\ \NON \\
&   &	\nopagebreak
\hskip 1.0in \hbox{ x }
\int_{0}^{\infty} d\veps {\pdr\over\pdr\veps}([1-f(\veps,s-s^\prime)]^2)
\eeqs
The $\veps$ integral can be trivially done. At the upper limit
$\veps = \infty , [1-f(\veps,s-s^\prime)]^2 = 1$ and the lower limit
$\veps = 0, [1-f(\veps,s-s^\prime)]^2 = 0$ which can be easily verified
using the definition of $f(\veps,s-s^\prime)$ in (26). Also the remaining
$s$ and $s^\prime$ integrals can be trivially done since in the limit
$m_0\rightarrow\infty$ we get
\beq
 m_0^2\hbox{sech}^2(m_0s)\hbox{sech}^2(m_0s^\prime)
= \ 4\delta(s)\ \delta(s^\prime)
\eeq
Using these results we get
\beq
\delta \cs_{eff}^{\rm chiral}=  -{i\over 4\pi}\int d^2z
			\Theta(z)[\eps^{ab}\pdr_a A_b(z)]
\eeq
Hence using (33) and (39) we finally get
\beq
\delta \cs_{eff}^a = \delta \cs_{eff}^{\rm cs} +
\delta \cs_{eff}^{\rm chiral} = 0
\eeq
\vspace{3ex}

\section{Gauge Invariance in terms of Currents}

	Having shown the cancellation of the gauge variation of the 1-loop
effective action we can try to look at the same phenomena in terms of the
currents. If we define the current as $J^\mu = \delta \cs_{eff}/\delta A_\mu$
we see from (30), after some simplification, that the current also can be
thought of as consisting of two components, Chern-Simons and chiral parts,
\beq
J = J^{\rm cs} + J^{\rm chiral}
\eeq
where we get
\beq
J^{\rm cs}_\mu = -{i\over 4\pi}{\rm sgn}(s)\eps_{\mu\rho\nu}\pdr^\rho A^\nu
			+ {i\over 4\pi}\delta_{\mu a}\eps_{ab}A^b \delta(s)
\eeq
and
\beq
\pdr^\mu J^{\rm chiral}_\mu = {i\over 4\pi}\eps_{ab}\pdr^a A^b \delta(s)
\eeq
which is the {\it consistent} anomaly in 2 dimensions as discussed in
 \cite{bard} and \cite{nacul}.
In the second term in (42) and in (43) the values of {\it a} and {\it b}
go over only 0,1. Now consider the second term in (42). This term is nonzero
only on the wall which suggests that
it be  considered with $J^{\rm chiral}$ which is nonzero only on the wall.
Then we obtain the following splitting of the currents.
\beq
J = J^{\rm GW} + J^{\rm cov}
\eeq
where
\beq
\left.
\matrix{
J^{\rm GW}_\mu = -{i\over 4\pi}{\rm sgn}(s)\eps_{\mu\rho\nu}\pdr^\rho A^\nu
\hbox{ and }
\cr
\cr
J^{\rm cov}_\mu = J^{\rm chiral}_\mu + {i\over 4\pi}\delta_{\mu a}
	\eps_{ab}A^b \delta(s)
\cr
\cr
}
\right.
\eeq
Clearly $J^{\rm cov}$ gives the covariant anomaly in 2 dimensions, which
is twice the consistent anomaly as it should be. As seen in (45), in addition
to the usual
chiral current, an extra term is needed to make the anomaly covariant. This
extra piece comes from the Chern-Simons piece as noted in \cite{nacul}.
$J^{\rm GW}$ on the other hand can be calculated using the methods of
Goldstone and Wilczek \cite{gold} by considering points far from the wall
where the approximation required for such an analysis is valid.

\section{Conclusion}

	We have shown explicitly that the potentially anomalous contributions
in the gauge variation of the effective action cancel between the Chern-Simons
term and the Chiral term. This was suggested in \cite{call} and
\cite{nacul}, but here we show that their analysis can be made exact even
close to the wall. We find the Chern-Simons term with the space dependent
coefficient as suggested before valid exactly even close to the wall. Further
we find that the effect of the massive modes is not only to produce
the Chern-Simons term but also to produce a chiral term in the propagator which
plays a critical role in the cancellation of the gauge variation. A closer
look indicates that the effect of the massive modes is to make the
ultra violet behavior softer for the chiral modes, acting as a regulator.
Further investigations of additional effects of
massive modes on the wall might be interesting in the context of \cite{kap},
where the effects of
the massive modes on the two dimensional domain wall is crucial. Also the claim
that the anomaly on the two dimensional wall is a covariant anomaly
can be better understood from this explicit calculation. The expectation of
Naculich \cite{nacul} appears to have been borne out by our calculation.
\vspace{3ex}

\noindent{\bf Acknowledgement}
\vspace{1ex}

	I would like to thank Prof. Norman Christ for his time, advise,
encouragement and lots of discussions throughout the period of this work. I
also would like to thank him for the critical reading of the manuscript.
I would also like to thank Prof. V P Nair for many discussions and helpful
suggestions. In the end I would like to thank Prof. Robert Mawhinney,
Dr. Zhu Yang and Prof. S G Rajeev for helpful comments.

\end{document}